\documentclass[referee]{0089} 
\usepackage{graphicx}
\usepackage{txfonts}
\newcommand{\kms}{km\,s$^{-1}$}
\newcommand{\HI}{H\,{\sc {i}}~}

\begin{document}
	\title{\HI and CO in the circumstellar 
	environment of the oxygen-rich AGB star RX Lep}

   \author{Y. Libert\inst{1}, 
	   T. Le Bertre\inst{1}, 
	   E. G\'{e}rard\inst{2},
	   \and J.M. Winters\inst{3}
          }

   \institute{LERMA, UMR 8112, Observatoire de Paris, 
	      61 Av. de l'Observatoire, 75014 Paris, France\\
              \email{yannick.libert@obspm.fr}
	\and
	      GEPI, UMR 8111, Observatoire de Paris, 
	      5 Place J. Janssen, 92195 Meudon Cedex, France
         \and
             IRAM, 300 rue de la Piscine, 
	     38406 St. Martin d'H\`{e}res, France
             }

\date{Received: 29/04/2008; accepted: 25/08/2008}
 
  \abstract
   {Circumstellar shells around AGB stars are built over long periods 
    of time that may reach several million years. They may therefore be 
    extended over large sizes ($\sim$ 1 pc, possibly more), and different 
    complementary tracers are needed to describe their global properties.}
   {We set up a program to explore the properties of 
   matter in the external parts of circumstellar shells around AGB stars 
   and to relate them to those of the central sources (inner shells and 
   stellar atmospheres).}
   {In the present work, we combined 21-cm \HI and CO rotational 
   line data obtained
    on an oxygen-rich semi-regular variable, RX Lep, to describe the global
    properties of its circumstellar environment.}
   {With the SEST, we detected the CO(2-1) rotational line from RX Lep.
    The line profile is parabolic and implies an expansion velocity of 
   $\sim$ 4.2 \kms ~and 
   a mass-loss rate $\sim$ 1.7\,$\times$\,10$^{-7}$\,M$_{\odot}$\,yr$^{-1}$ 
   (d $=$ 137 pc). The \HI line at 21 cm was detected with the Nan\c{c}ay 
   Radiotelescope on the star position and at several offset positions. 
   The linear shell size is relatively small, $\sim$ 0.1 pc, but 
   we detect a trail extending southward to $\sim$ 0.5 pc. 
   The line profiles are approximately Gaussian with an FWHM $\sim$ 3.8 \kms 
   ~and interpreted with a model developed for the detached shell around the 
   carbon-rich AGB star Y CVn. Our \HI spectra are well-reproduced by assuming a 
   constant outflow ($\rm \dot{M}$ $=$ 1.65 $\times$ 10$^{-7}$ M$_{\odot}$ 
   yr$^{-1}$) of $\sim$ 4 $\times$ 10$^{4}$ years duration, which has been 
   slowed down by the external medium. 
   The spatial offset of the \HI source is consistent with the northward 
   direction of the proper motion
    measured by Hipparcos, lending support to the presence of a trail resulting
    from the motion of the source through the ISM, as already suggested for 
   Mira, RS Cnc, and other sources detected in \HI. The source was also observed 
   in SiO (3 mm) and OH (18 cm), but not detected.
   }
   {A detached shell, similar to the one around Y CVn, was discovered in \HI around
    RX Lep. We also found evidence of an extension in the direction opposite to 
   the star proper motion. The properties of the external parts of circumstellar shells 
   around AGB stars should be dominated by the interaction between stellar 
   outflows and external matter for oxygen-rich, as well as for carbon-rich, 
   sources, and the 21-cm \HI line provides a very useful tracer of these 
   regions.}


   \authorrunning{Libert et al.}
   \titlerunning{\HI \& CO in RX Lep.}
   \maketitle
%

\section{Introduction}
\label{intro}
	Evolved stars on the asymptotic giant branch (AGB) 
	are often surrounded by circumstellar shells. 
	The material in these shells is flowing outwards with 
	velocities from a few~\kms~up to 40 km s$^{-1}$ (Nyman et al. 1992). 
	The observed mass-loss rates range from $\sim$ 10$^{-8}$ to a few 
	10$^{-4}$ M$_{\odot}$ yr$^{-1}$ (e.g. Knapp \& Morris 1985; 
	Olofsson et al. 2002), the lower limit being probably set by 
	detectability. In this phase of the stellar life, the evolution is 
	dominated by mass loss rather than nuclear processes (Olofsson 1999). 
	The history of mass loss over the full AGB is complex and the details of this
	process are currently not well known (e.g., Lafon \& Berruyer 1991;
	Habing 1996; Le{\~a}o et al. 2006). A general picture, however, has arisen
	from both theoretical and observational findings that - on
	average - the mass-loss rate increases towards the end of the AGB
	phase, leading in some cases to the formation of a planetary nebula
	(e.g., Renzini 1981; Hrivnak \& Bieging 2005).

	The validity of this simple picture may depend on 
	the parameters of the star, e.g., on its initial mass. Schr{\"o}der 
	et al. (1999) combine mass-loss rates derived from consistent wind models 
	with stellar evolution calculations and find that the mass-loss rate
	should increase along the AGB for stars with 
	initial masses greater than 1.3 M$_{\odot}$. 
	Stars with lower initial mass would experience a single
	short-lived ($\sim$ 1000 yr) 
	episode of high mass loss only, which would leave behind a
	very narrow detached shell as observed in the case of, e.g., TT Cyg
	(Olofsson et al. 2000). On the other hand, the mass-loss phenomenon
	appears to be highly variable on even shorter time scales as 
	indicated e.g. by concentric arcs observed in scattered light around
	the prototype carbon Mira IRC~+10216 (Mauron \& Huggins 2000) and
	around some proto-planetary nebulae 
	(e.g., Hrivnak et al. 2001). The time scale of these
	mass-loss variations would be around a few 10$^2$\,yr. The
	physical mechanism responsible for these variations still needs to be
	identified, although different possibilities have already been
	proposed: e.g., interaction between gas and dust within stellar outflows 
	(Simis et al. 2001), or solar-like magnetic cycle (Soker 2002). In 
	contrast to these later phases of AGB mass loss at rather high
	rates ($\sim$ 10$^{-5}\rm M_{\odot}$ yr$^{-1}$), information about the mass-loss
	process on the early AGB, is even scarcer.

	To unravel the processes involved in the mass-loss phenomenon, 
	we have to find suitable tracers. One of the most studied among 
	these tracers is the CO molecule, because so far it has been considered 
	to provide the best estimate of the mass-loss rate for AGB stars 
	(Ramstedt et al. 2008).
	Not only can it be used to estimate this mass-loss rate, but it also 
	yields important parameters of the AGB wind (e.g.: expansion 
	velocity, central star velocity, etc.). Since CO is photodissociated by 
	UV radiation from the 
	interstellar radiation field (ISRF), it can only probe the 
	inner parts (r $\leq$ 10$^{-3}$ - 10$^{-1}$ pc, Mamon et al. 1988) 
	of circumstellar 
	shells (CSs). Therefore, the CO emission is only related to ``recent'' 
	(i.e. a few 10$^{3}$ - 10$^4$ years) mass-loss episodes. 

	On the other hand, \HI is in general protected from photoionization 
	by the surrounding interstellar medium (ISM). As a result, \HI can be 
	used to probe the external parts of circumstellar shells and can give 
	indications on the mass-loss on longer timescales (a few 10$^5$ 
	years: Libert et al. 2007). Hence, CO and \HI complement each 
	other nicely to describe the history of the mass-loss rate of an AGB star.

	The drawbacks of \HI circumstellar observations are that hydrogen is 
	ubiquitous in the Galaxy and that the genuine stellar \HI must be 
	separated from the ambient H\,{\sc {i}}. Ideal cases would be bright \HI 
	sources, with relatively high velocity with respect to the local 
	standard of rest (LSR) and reasonably far above the Galactic plane. 
	For the other sources, the interstellar \HI should be studied with 
	care. In the present paper, we analyze this confusion 
	with a new approach that consists in a 3D-mapping using \HI spectra.

	The mass-loss phenomenon is different from one AGB star to another 
	and may vary highly with time. Nevertheless, 
	observing in \HI provides a global view of the CS behavior and, on 
	timescales of about 10$^5$ years, small variations in the mass-loss 
	rate may be flattened out. Thus, we have developed a model of the 
	circumstellar gas, based on a scenario already proposed by Young 
	et al. (1993), in which CSs are the result of a constant outflow 
	eventually slowed down by the surrounding medium. This deceleration 
	produces a snowplough effect around the source, resulting in a 
	detached shell of compressed matter originating from the 
	star and the external medium (Lamers \& Cassinelli 1999, Chap. 12).

	A schematic view of this model can be pictured as follows: a wind 
	is flowing outward from the star, in free expansion with a constant 
	velocity (V$\rm _{exp}$) and a constant rate. It encounters a shock 
	at a radius r$_1$ (termination shock), due to the slowing down by the 
	surrounding matter. Between r$_1$ and r$\rm _f$ (contact 
	discontinuity), the stellar matter 
	is compressed. Between r$\rm _f$ and until a second shock at r$_2$ 
	(bow shock), the interstellar matter has been swept up by the wind 
	of the AGB star. Finally, beyond r$_2$, the external matter is 
	considered to be at rest.

	Recently, we successfully applied this model to a carbon-rich star: Y CVn 
	(Libert et al. 2007). In \HI at 21 cm, this star exhibits a composite profile, 
	made of a broad, rectangular component and a narrow, Gaussian-shaped 
	one. In our description, the broad component is the signature of the 
	freely expanding wind, whereas the narrow component is produced by 
	the \HI compressed in the snowplough between r$_1$ and r$_2$. Our 
	model provides a simple explanation for some of the so-called 
	``detached dust shells'' observed in the far infrared (Izumiura 
	et al. 1996). If this approach is correct, then it should also apply 
	to detached shells around oxygen-rich AGB stars.
	In this paper we present \HI and CO data that we obtained 
	on an oxygen-rich AGB star, RX Lep, and interpret them with the 
	model that we developed for Y CVn. The \HI interstellar 
	confusion in the direction of RX Lep is moderate and we illustrate, 
	in that case, our new approach to extract a genuine \HI spectrum.

	In this simplified description we assume spherical symmetry. However,
	recent H\,{\sc {i}}, far-infrared and UV data (G\'erard \& Le\,Bertre 2006; Matthews
	\& Reid 2007; Ueta et al. 2006; Martin et al. 2007) have shown that
	the AGB star motion with respect to the ISM may lead to a distortion,
	and eventually a disruption, of the circumstellar environment.
	Previous, and more recent, numerical modelings (Villaver et al. 2003;
	Wareing et al. 2007) are in line with this interpretation.
	The circumstellar environment of RX Lep might provide a new illustration
	of this phenomenon, and we will discuss this possibility.

\section{RX Lep}
\label{source}
	RX Lep has been classified as an irregular variable, Lb star 
	(General Catalogue of Variable Stars, GCVS 3rd ed., Kukarkin 
	et al. 1971). A photometric monitoring over 8 years shows 
	variations of about $\pm$ 1 magnitude in the V band (Cristian 
	et al. 1995). The periodogram analysis gives a main period in 
	the range 80-100 days and a possible secondary period around 
	60 days. Recently, the star has been re-assigned to the type 
	SRb (GCVS 4.2, Samus et al. 2004), because it may exhibit a periodic 
	variability of a few tenths of a magnitude.

	The Hipparcos parallax (7.30$\pm$0.71 mas) places the star at 
	137$^{+15}_{-12}$ pc from the Sun and at 
	$\sim$\,65\,pc away from the Galactic plane 
	(b$^{\rm II}$$=$-27.51$^{\circ}$). The proper motion, also given 
	by Hipparcos, is 31.76$\pm$0.58 mas\,yr$^{-1}$ in right ascension (RA) 
	and 56.93$\pm$0.50 mas\,yr$^{-1}$ in declination (Dec.). At 137 pc, it 
	translates into a motion in the plane of the sky of 44 
	km s$^{-1}$ (corrected for solar motion, as determined by Dehnen 
	\& Binney 1998) in the northeast direction (PA $\sim$ 31$^{\circ}$).

	Fouqu\'{e} et al. (1992) have obtained near-infrared photometry 
	data and, using the bolometric correction of Le Bertre et al. 
	(2001), we derived a luminosity of 4500 L$_{\odot}$. This 
	luminosity confirms that RX Lep is on the AGB. The effective 
	temperature is $\sim$ 3300 K (Dumm \& Schild 1998). This means 
	that hydrogen is expected to already be mostly in atomic form in 
	the atmosphere and throughout the CS (Glassgold \& Huggins 1983). 
	Technetium lines ($^{99}$Tc) were searched for in the 4200-4300 
	\AA ~region and not detected (Lebzelter \& Hron 1999), confirming 
	an older result from Little et al. (1987). This tends to indicate 
	that RX Lep has not gone through a thermal pulse and that it is 
	still in the early phase of the AGB (E-AGB). This is in good 
	agreement with the results of Mennessier et al. (2001) who, using 
	astrometric and kinematic data, place RX Lep among E-AGB stars 
	that belong to the Galactic disk population with initial masses in 
	the range 2.5-4 M$_{\odot}$.

	Paschenko et al. (1971) did not detect the source in the OH satellite 
	line at 1612 MHz. As this is the only OH observation reported in the 
	literature, we observed RX Lep again at 18 cm on Jan. 12, 2006 and July 
	14, 2006 with the NRT. No emission was detected at a level of 0.015 
	Jy in any of the 4 OH lines (1612, 1665, 1667, and 1720 MHz).

	RX Lep might be associated with an IRAS extended source 
	(X0509-119, IRAS Science 
	Team 1988) at 60 (diameter $\sim$ 1.1$'$) and 100 $\mu$m (diameter 
	$\sim$ 6.0$'$). However, X0509-119 is centered at about 2.5$'$ east 
	from RX Lep, and that association might only be fortuitous. We 
	present a re-analysis of the IRAS results farther down (Sect.~\ref{discussions}).

	Kerschbaum \& Olofsson (1999) report a CO (1-0) and CO (2-1) 
	detection, but their radial velocity is doubtful:  
	v$\rm _{hel}$\,$\sim$\,29\,km\,s$^{-1}$ (as compared to 
	v$\rm _{hel}$\,$\sim$\,46\,km\,s$^{-1}$ cited in the General Catalogue of 
	Stellar Radial Velocities, GCRV, Wilson 1953). Our new results 
	(Sect.~\ref{CO}) now suggest there has likely been a confusion between the 
	heliocentric and LSR reference frames.

\section{Molecular line observations}
\label{CO}
	RX Lep was part of a CO observing program dedicated to the 
	Valinhos 'b' class stars (Epchtein et al. 1987). This class of sources 
	is defined by a weak near-IR excess as compared with the IRAS fluxes 
	(0.2 $<$ K-L' $<$ 0.7 and 0.8 $<$ L'-[12] $<$ 2). The central stars are 
	generally identified with late-M giants surrounded by tenuous circumstellar shells. 
	Those stars were suspected by Winters 
	et al. (2000) to show preferentially low expansion velocity winds. 
	Most of the data from this program  have been published in 
	Winters et al. (2003). Subsequently, RX Lep's CO (2-1) emission at 
	230\,GHz and SiO (v=1, J=2-1) maser transition at 86\,GHz have been 
	searched using the 15-m Swedish-ESO Submillimetre Telescope, SEST 
	(Booth et al. 1989) on January 30, 2003. At 1.3\,mm, the FWHM of 
	the SEST beam is 23$''$. We used the position-switch mode with a beam 
	throw of 11.5$'$. The spectra were recorded on the high-resolution 
	spectrometer (HRS) giving a resolution of 80 kHz, for a channel 
	separation of 43 kHz and a bandwidth of 86 MHz. 

\begin{figure}
\centering
\includegraphics[width=9cm]{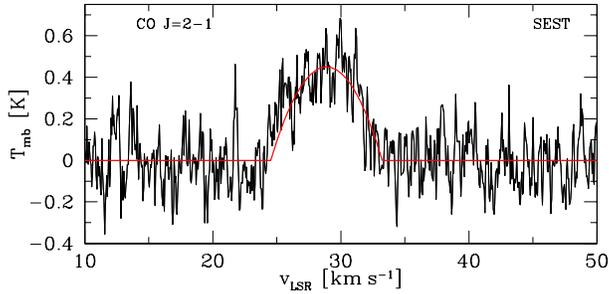}
\caption{CO (2-1) emission of RX Lep. The solid line represents a parabolic 
	fit (Sect.~\ref{CO}).}
\label{cofit}
\end{figure}

	The CO (2-1) transition was clearly detected (Fig.~\ref{cofit}). 
	The resulting profile was fitted with a parabolic curve, and we 
	derived an LSR velocity of V$\rm _{lsr}$ $=$ 28.9 $\pm$ 0.1 
	km s$^{-1}$, an expansion velocity of V$\rm _{exp}$ $=$ 4.2 
	$\pm$ 0.1 km s$^{-1}$, and an amplitude of T$\rm _{mb}$\,$=$\,0.45\,K 
	$\pm$ 0.13 K.
	By using the same method as Winters et al. (2003, Sect. 4.3), we 
	estimated both RX Lep mass-loss rate and CO photo-dissociation radius 
	using the results of the line fitting. We find $\rm \dot{M}$ $\sim$ 
	1.7 $\times$ 10 $^{-7}$ M$_{\odot}$ yr$^{-1}$ and r$\rm _{CO}$ 
	$\sim$ 0.8 $\times$ 10$^{-2}$ pc ($\equiv$\,12.5$''$). On the other 
	hand, the SiO maser was not detected at a level of 0.2\,Jy.

	Our CO measurement of the LSR radial velocity (28.9~\kms) is consistent with 
	the heliocentric velocity quoted in the GCRV. Combining this result with 
	the Hipparcos determination of the velocity in the plane of the sky (44~\kms), 
	we get a 3-D space velocity of 53~\kms.

\section{\HI observations}
\label{HI}

\begin{figure}
\includegraphics[width=5cm]{0089Fi2a.ps}\\
\includegraphics[width=5cm, angle=-90]{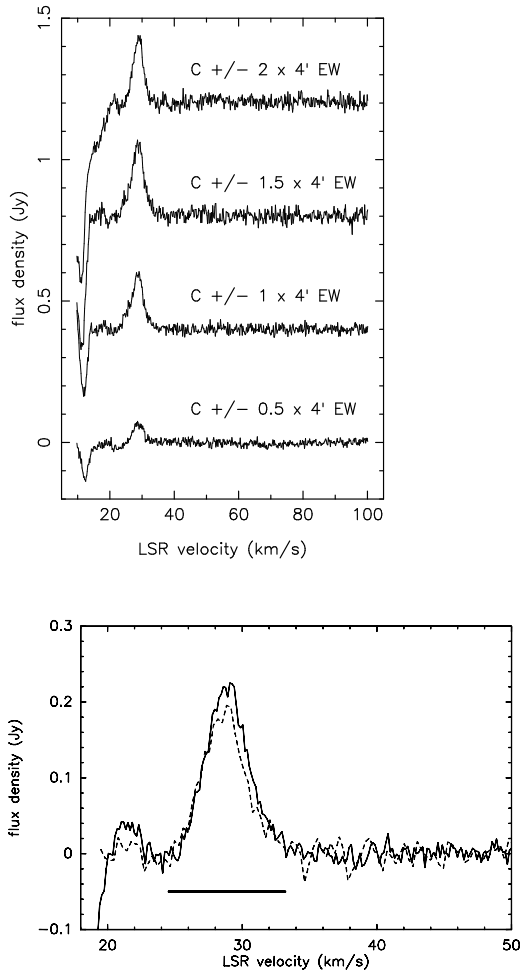}
\caption{Upper panel: Spectra obtained in position-switch mode with the NRT. 
	The positions are expressed in number of beams (4$'$). 
	The maximum intensity is reached at C$\pm$1.5\,$\times$\,4$'$\,EW. 
	For clarity, the individual spectra have been displayed 
	with vertical offsets of 0.4 Jy.
	Lower panel: average spectrum computed with 
	C\,$\pm$\,n\,$\times$\,4$'$\,EW, n $>$ 1.5. Dotted line: baseline 
	subtracted f-switch 
	spectrum. The horizontal line shows the width of the CO signal.}
\label{offset}
\end{figure}

\begin{figure*}
\includegraphics[width=9cm]{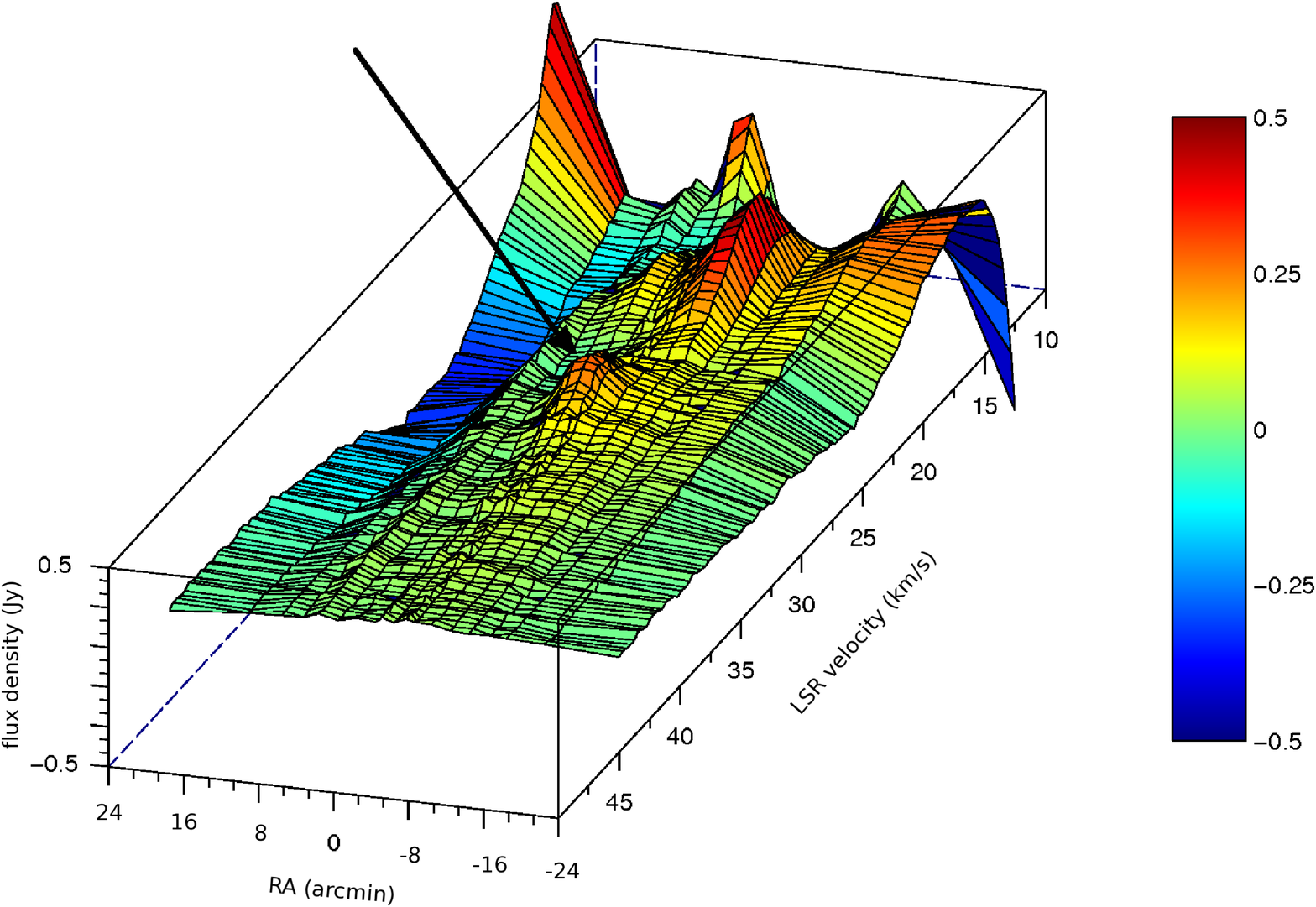}
\includegraphics[width=9cm]{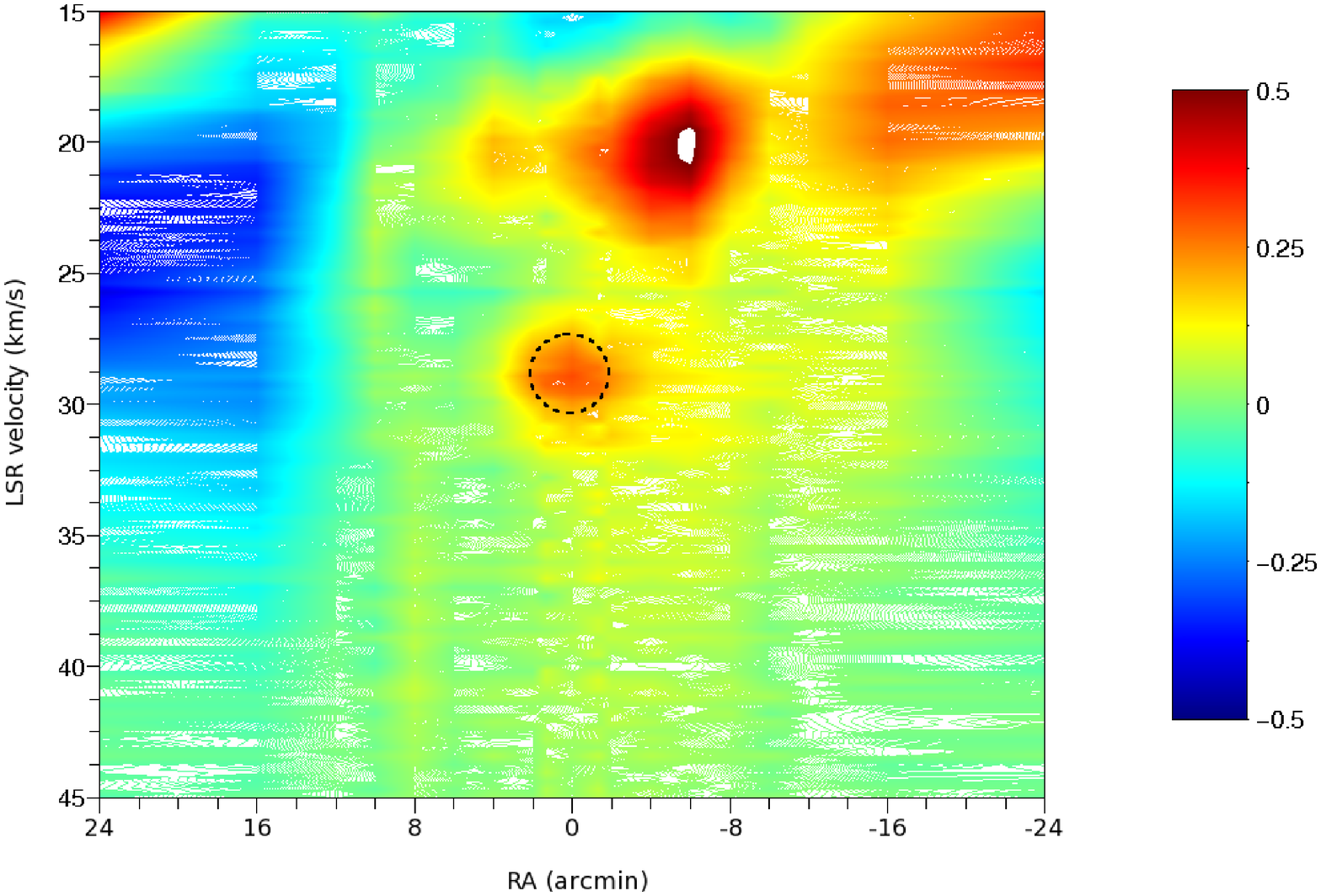}
\caption{Left panel: 3D velocity-position representation of the \HI flux 
	density; east is to the left. The arrow points to 
	the expected position of the source. Right panel: the same data set 
	represented in 2D; west is to the right. The dashed circle surrounds 
	the expected position of RX Lep.}
\label{hImap1}
\end{figure*}

\begin{figure*}
\centering
\includegraphics[width=10cm, angle=-90]{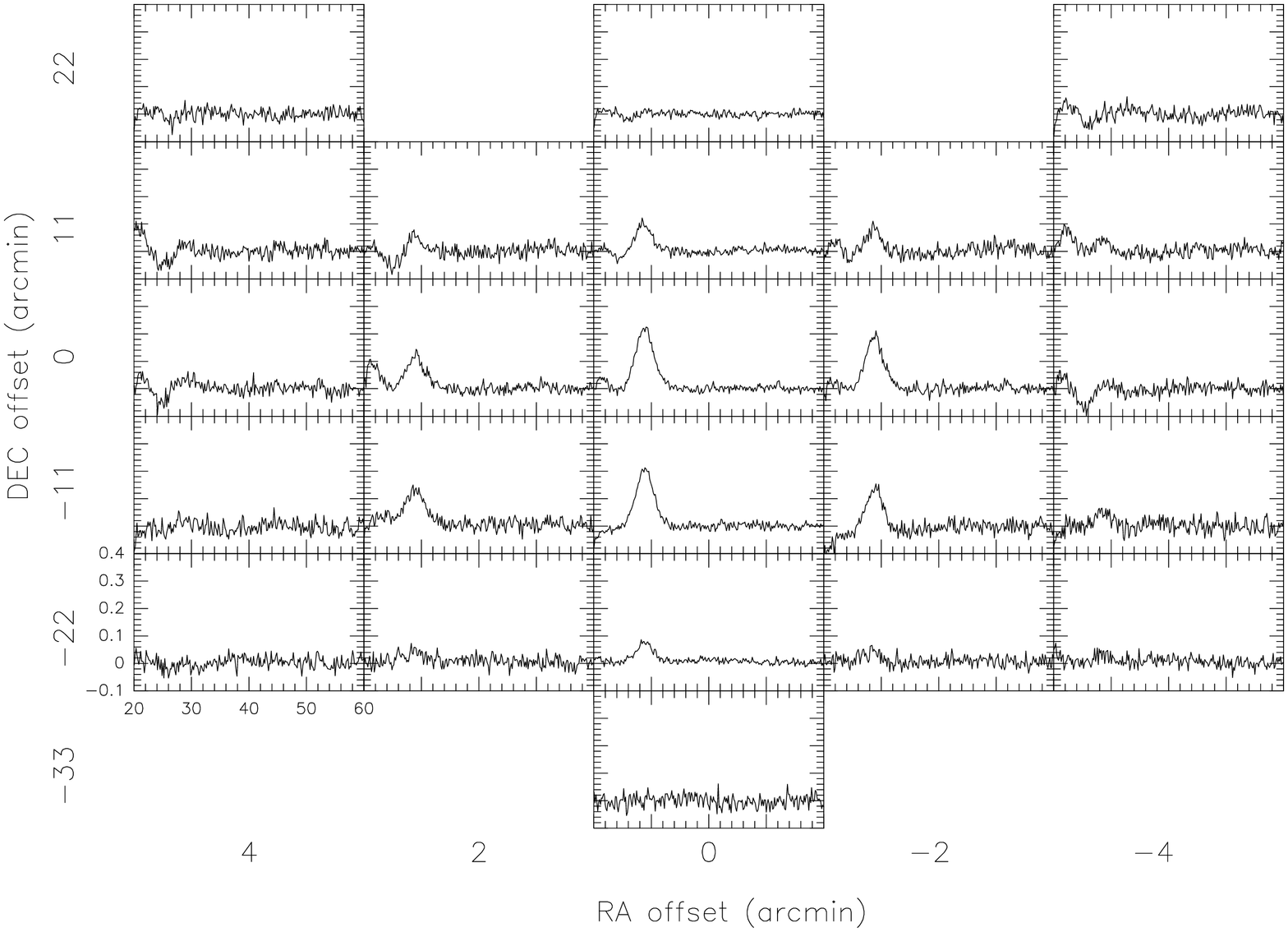}
\caption{\HI map of RX Lep. The steps are 2$'$ in RA and 11$'$ in Dec. 
	The positions are indicated with respect to the stellar position. 
	The abscissae and ordinates are LSR radial velocities (\kms) and 
	flux densities (Jy) as indicated on the lower left corner.}
\label{hImap2}
\end{figure*}

	RX Lep has been observed during a total of 141 hours 
	between February 2005 and February 2008 with the Nan\c{c}ay 
	Radiotelescope (NRT). The NRT is a meridian telescope with a 
	rectangular aperture of effective dimensions 160$\times$30 m. At 21 
	cm and at the declination of RX Lep, the FWHM of the beam is 4$'$ 
	in RA and 22$'$ in declination. We used the position-switch 
	technique with two off-positions in the east-west direction, every 2$'$, 
	and up to 24$'$ from the source. Thus, the total time 
	spent on-source was 47 hours. To fully describe the environment of 
	RX Lep, we sampled our map every half beam in RA and in 
	Dec. (hereafter, the position-switch spectra will be referred 
	to as C\,$\pm$\,n\,$\times$\,4$'$\,EW, n being the number of beams for 
	the off positions). 
	At 21 cm, the spectra have a bandwidth of 165 km 
	s$^{-1}$ and a channel width of 0.08 km s$^{-1}$. For convenient 
	analysis, we smoothed the data with a Hanning filter so that the 
	spectral resolution was 0.16 km s$^{-1}$. The data are processed with 
	the CLASS software, part of the 
	GILDAS\footnote{http://www.iram.fr/IRAMFR/GILDAS} 
	package developed at IRAM (Pety 2005).

	The different steps of the data processing can be described as 
	follows. First of all, we determine the spatial extent of the 
	source by comparing the C\,$\pm$\,n\,$\times$\,4$'$\,EW spectra. When the maximum 
	intensity of the peak is reached (n = n$\rm _{max}$), the source does 
	not contribute to the flux of the offset spectra anymore. For 
	example, according to Fig.~\ref{offset} (upper panel), RX Lep does not extend 
	farther than 6$'$ in the E-W direction. Once the maximum extent 
	is estimated, the average of the C\,$\pm$\,n\,$\times$\,4$'$\,EW spectra with 
	n $>$ n$\rm _{max}$ gives the central spectrum (Fig.~\ref{offset}, lower panel). 
	Simple arithmetic then allows to extract the spectra at the offset 
	positions, using the central spectrum. 

	We present a new visualization of the \HI spectra to better separate 
	the genuine stellar \HI from the contamination due to interstellar 
	hydrogen. The operation can be described as a stacking of the NRT 
	spectra, processed as above, in the east-west direction, for a given 
	declination (Fig.~\ref{hImap1}). In this view, velocity is given as 
	a function of right ascension, and intensity is represented using a 
	colored scale. This aims at visualizing, and thus separating, the \HI 
	emission coming from the source and that from the Galaxy. 
	Indeed, on the resulting image, the stellar \HI should be nearly 
	centered in RA and close to the LSR velocity given by CO observations. 
	While this process emphasizes the difficulties coming from the 
	contamination due to the Galactic hydrogen emission, it also allows 
	an evaluation of the possible problems when processing the data and a 
	design of the best strategy for extracting the intrinsic source emission.

 	According to Fig.~\ref{hImap1}, RX Lep is definitely a suitable 
	candidate for \HI observation, as it is clearly 
	separated from the interstellar emission spectrally and 
	spatially, although the confusion increases 
	for velocities lower than 29 km s$^{-1}$. Indeed, the image shows 
	2 potential sources of contamination: one around 4$'$W from the source 
	and at $\sim$ 20~\kms , the other increasing (negatively) at 16$'$E and 
	around 24~\kms . 
	This information is crucial to safely extracting the intrinsic 
	emission of RX Lep. Thus, to build the map of the source, 
	we fitted polynomial baselines (in some cases of degree up to 3 
	when the confusion reaches its highest level) to a portion of the 
	spectrum between 21 and 54\,km\,s$^{-1}$. The resulting map of RX Lep 
	is shown in Fig.~\ref{hImap2}. 

	We independently confirmed these results by 
	observing the source using the frequency-switch mode 
	(Fig.~\ref{offset}, lower panel). We spent 5 hours on source 
	and detected it at the same velocity and with the same flux density 
	as shown on the map for the central position.

	From our observations, we can readily derive some important 
	properties of the CS. The map of RX Lep in Fig.~\ref{hImap2} 
	reveals that the \HI line profile, at the central position, shows a 
	quasi-Gaussian shape of central velocity 28.84$\pm$0.03 
	km s$^{-1}$, FWHM 3.8$\pm$0.1 km s$^{-1}$, and flux density 
	0.22$\pm$0.03\,Jy. The shape of this line differs from that of the 
	parabolic CO line. It clearly indicates a slowing down of the wind 
	within the outer parts of the CS (Le\,Bertre \& G\'{e}rard 2004). 
	Moreover, assuming that the broadening of the \HI emission line is 
	dominated by the thermal Doppler effect, the FWHM of the spectrum 
	allows us to estimate an upper limit to the average temperature in the 
	shell (Libert et al. 2007, Eq. 1). It gives us T$\rm _{mean}$ $<$ 
	312 K.

\begin{figure}[b]
\centering
\includegraphics[width=5cm, angle=-90]{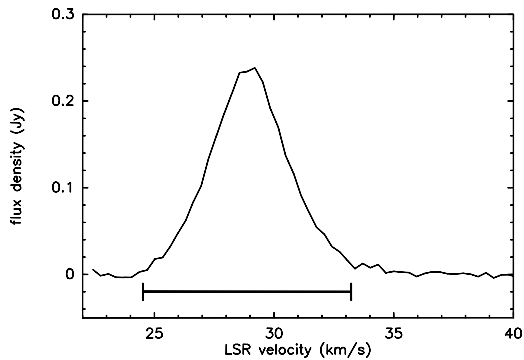}
\caption{ Average of the spectra C $\pm$ n $\times$ 4$'$ EW and 11$'$S 
	$\pm$ n $\times$ 4$'$ EW with n $>$ 1. The horizontal line represents the 
	width of the CO line. }
\label{pied}
\end{figure}

	Evidence of a composite line profile such as that of Y CVn 
	(Sect.~\ref{intro}) is difficult to see, given the fairly low 
	intensity of the signal and the narrow width of the expected pedestal 
	(2 $\times$ V$\rm _{exp}$). Nevertheless, from the 
	central spectrum, we can set an upper limit to the amplitude of a 
	possible pedestal. We estimate this limit at 20 mJy by assuming it has 
	the same width as the CO profile (Fig.~\ref{pied}). 

	The map of RX Lep (Fig.~\ref{hImap2}) shows that the \HI brightness 
	distribution of the envelope is offset from the stellar position both 
	in RA and Dec. There is a slight westward RA offset $<$ 1$'$ (since the 
	flux density at 2$'$ west is larger than at 2$'$ east but smaller 
	than at the center). There is a southward Dec. offset close to 5.5$'$ 
	(since the flux density at 11$'$ south is nearly equal to the central 
	flux). It is useful to give quantitative estimates, not only of the 
	offsets but also of the spatial extents for the model calculations 
	that will be discussed in Sect.~\ref{model}. If one assumes that the \HI 
	brightness distribution is Gaussian in RA (and Dec.) and offset, 
	the convolution by a Gaussian beam also produces a Gaussian distribution 
	and one can retrieve from the data both the offset 
	and half power width (HPW) in 
	RA (and Dec.). The RA offset and HPW are respectively~ $-$0.4$'$ 
	($\pm$ 0.2$'$) and 2.3$'$ ($\pm$ 0.5$'$). The Dec. offset and HPW 
	are respectively~$-$4.4$'$ ($\pm$ 0.6$'$) and 15$'$ ($\pm$ 3$'$). 
	Thus the \HI envelope is elongated southwards and offset from the 
	stellar position by 4.4$'$ at PA 185$^{\circ}$ (i.e. also nearly 
	southward). This suggests an HI envelope trailing south.   

	The integrated flux throughout the map gives 1.22\,Jy$\times$\kms, which 
	translates into a hydrogen mass of $\sim$ 5.42\,$\times$\,10$^{-3}$\,
	M$_{\odot}$ (assuming no hydrogen in H$_2$; cf. Sect.~\ref{source}). 
	Adopting a mean molecular weight of 1.3, it translates 
	into a total mass of the gas in the shell of $\sim$ 7.05 $\times$ 10$^{-3}$ 
	M$_{\odot}$. If we consider the mass-loss rate to be constant and 
	adopt the estimate given by CO, then the age of the CS is $\sim$ 
	42700 years, about one order of magnitude less than the age we 
	estimated for Y CVn.

\section{Model}
\label{model}

	The high-quality \HI spectral profiles that we have obtained in the 
	direction of RX Lep are similar to those of Y CVn (Libert et al. 2007). This 
	type of profile is indicative of a slowing-down of stellar outflows in the 
	external parts of CSs (Le Bertre \& G\'{e}rard 2004). In the following we 
	apply the model that we developed for the carbon-rich star Y CVn in 
	order to evaluate the physical conditions within the shell of RX Lep. Of course, 
	as this model assumes sphericity, it cannot reproduce the more complex 
	geometry suggested by the map presented in the previous section. In the 
	east-west direction the map is fairly symmetric, so the model could apply. 
	However, there is also a clear north-south extension that would require a 2-D 
	model, as well as a spatial resolution better than 22$'$ (the NRT beam).
	
	A 1D-hydrodynamic 
	code provides the density distribution within the detached shell 
	based on the hypothesis of a slowing down of the stellar 
	gas by the surrounding local material. The mass-loss rate is constant, 
	and the gas expanding outward from the atmosphere is in free 
	expansion with a constant velocity V$\rm _{exp}$. Then, the outflow 
	encounters a shock (r$_1$). Its velocity decreases by a factor of 
	about 4 and the matter keeps on decelerating until it reaches the 
	external medium (at r$\rm _f$). The external matter that has been 
	swept up and compressed by the expansion of the stellar envelope 
	lies outside r$\rm _f$. Finally, beyond r$_2$, the gas is at rest.

	The expansion velocity and the LSR velocity of the source are based 
	on the results from our CO observations. But RX Lep has not been 
	studied much, so we lack some spatial information such as 
	estimates of r$_1$ and r$_2$ that could have been obtained, for 
	example, with dust continuum observations. Nevertheless, the NRT map 
	indicates that the object is fairly small in the east/west direction
	($\sim$ 2.3$'$ i.e. $\sim$ 9 $\times$ 10$^{-2}$ pc at 137 pc).

	One of the results of our model is that detached shells are flagged 
	by a composite \HI spectrum. The first component (Comp. 1) is narrow, with 
	a quasi-Gaussian shape and it arises from the matter slowed down by the 
	local medium. The second component (Comp. 2) is broad with a 
	rectangular shape, as it probes the inner part of the shell where 
	the gas is in free expansion. In Sect.~\ref{HI}, we set 
	an upper limit of Comp. 2 of $\sim$ 20 mJy. For a constant mass-loss 
	rate, we can derive a relation (Eq.~\ref{r1}) to estimate r$_1$:
	\begin{equation}
	r_1 \approx 2.17 \times 10^{-9} \times 
		\frac{d \ V_{exp}^2 \ F_{Comp.\ 2}}{\dot{M}} 
	\label{r1}
	\end{equation}
	where r$_1$ is expressed in arcmin, d is the distance in pc, 
	V$\rm _{exp}$ is in km s$^{-1}$, F$\rm _{Comp. 2}$ is the intensity 
	of the pedestal in Jy, and $\dot{\rm M}$ is in M$_{\odot}$ yr$^{-1}$. 
	With V$\rm _{exp}$ $=$ 4.2 km s$^{-1}$ and $\rm \dot{M}$ $=$\,
	1.65\,$\times$\,10$^{-7}$\,M$_{\odot}$\,yr$^{-1}$ (Sect.~\ref{CO}), we 
	estimate an upper limit for r$_1$ of 0.64$'$. We set r$_2$ at 
	1.15$'$, in agreement with the \HI observations in the 
	east-west direction (HPW/2, Sect.~\ref{HI}).

\begin{figure}
\includegraphics[width=6cm, angle=-90]{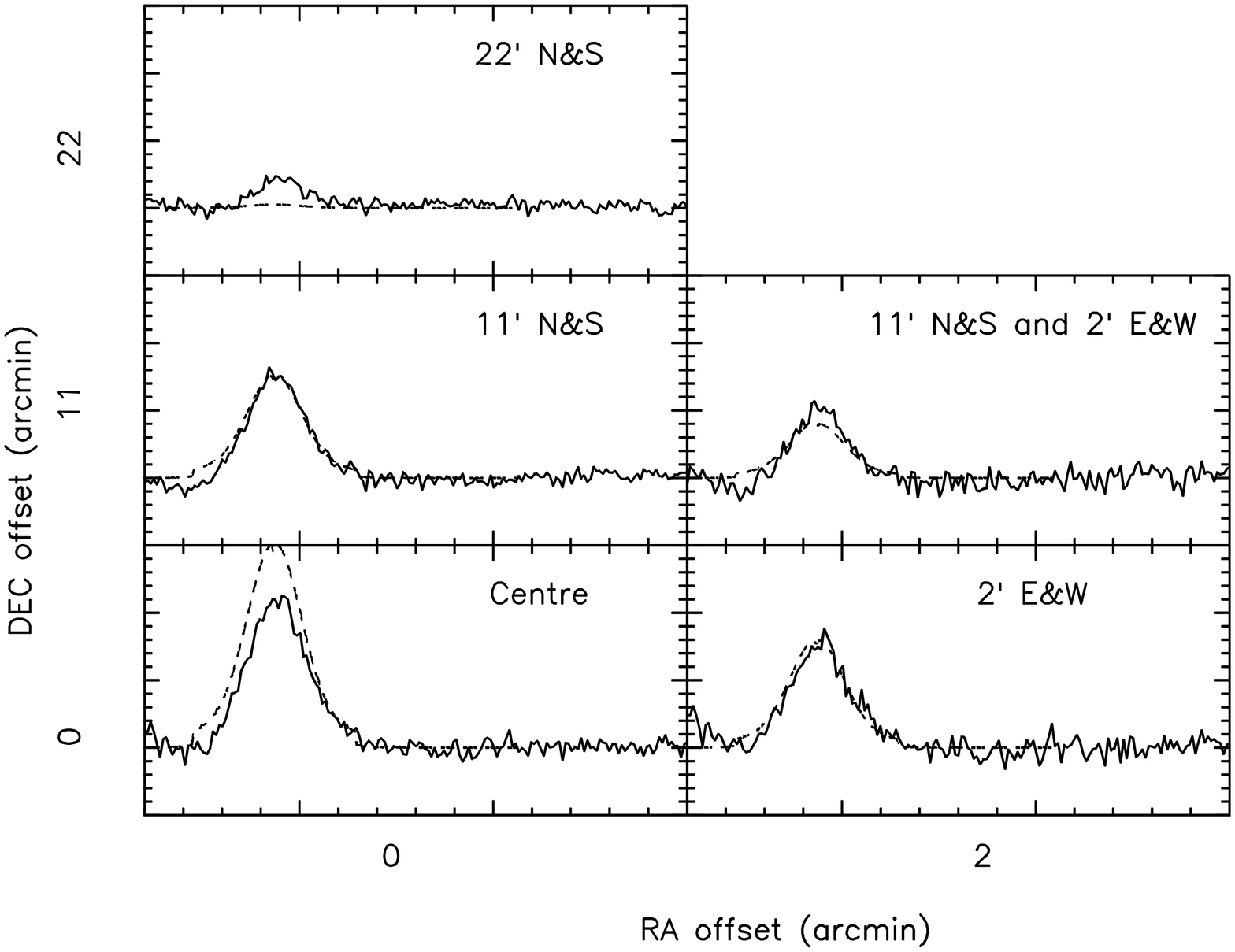}\\
\includegraphics[width=7cm, angle=-90]{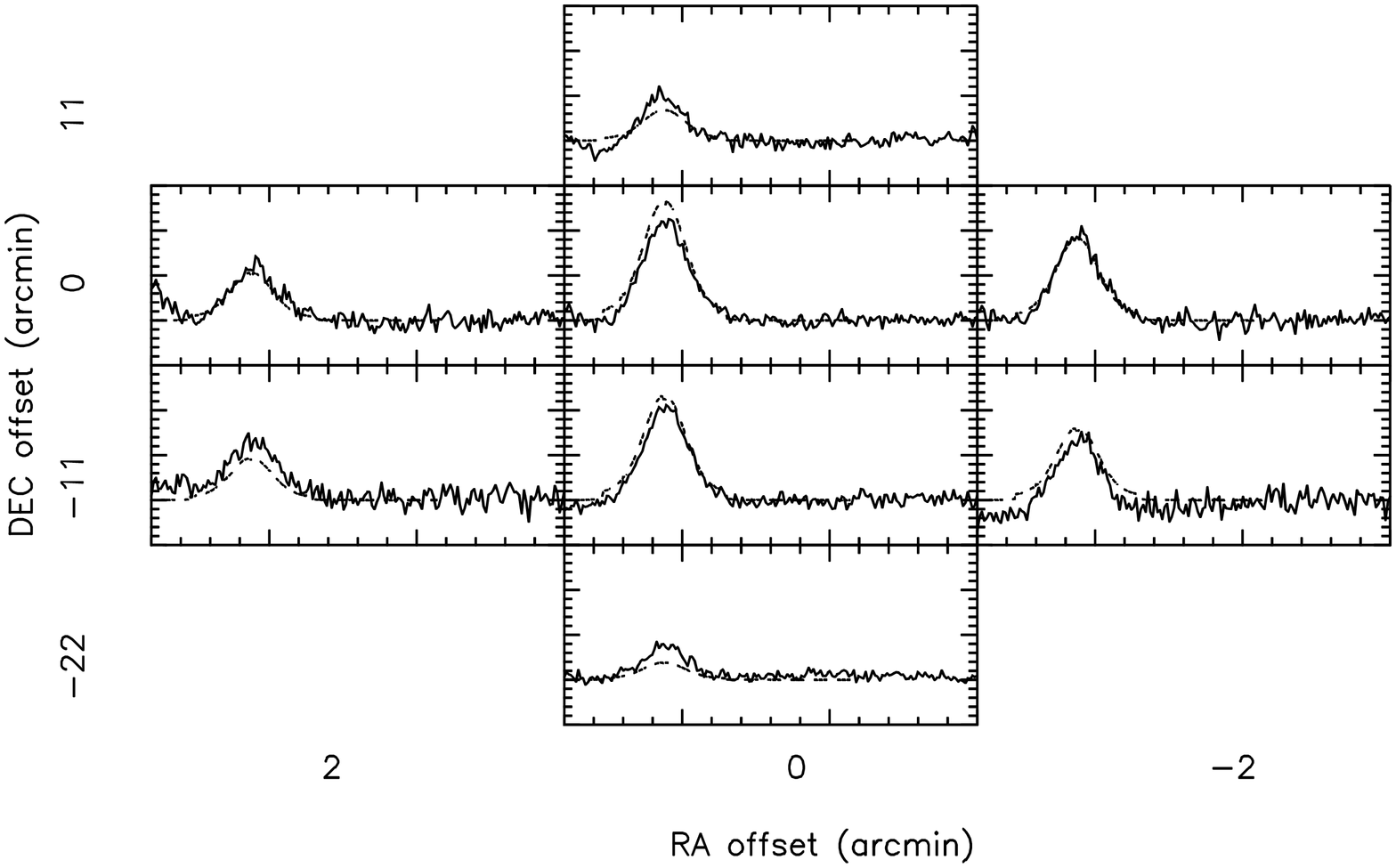}
\caption{\HI observations vs model (dashed line): the upper panel shows 
	the model discussed in Sect.~\ref{model} and compared to a 
	symmetrized \HI map of RX Lep. The lower panel presents the 
	same model shifted by 4.4$'$S and 0.4$'$W 
	and compared to the \HI map of RX Lep (as in Fig.~\ref{hImap2}).}
\label{model_HI}
\end{figure}

\begin{figure}
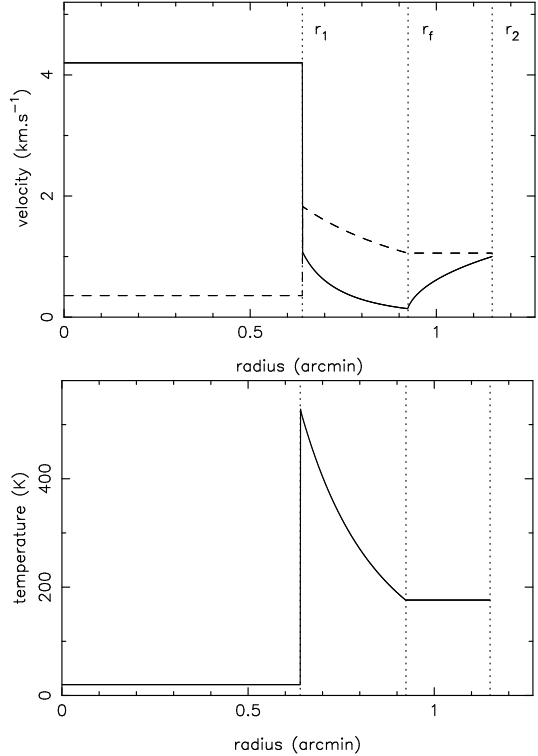

\centering
\includegraphics[width=5cm, angle=-90]{0089Fi7a.ps}\\
\includegraphics[width=5cm, angle=-90]{0089Fi7b.ps}
\caption{Upper panel: velocity profile. The dashed line represents the isothermal 
	sound velocity. Lower panel: temperature profile adopted for the model.}
\label{profvt}
\end{figure}

\begin{figure*}
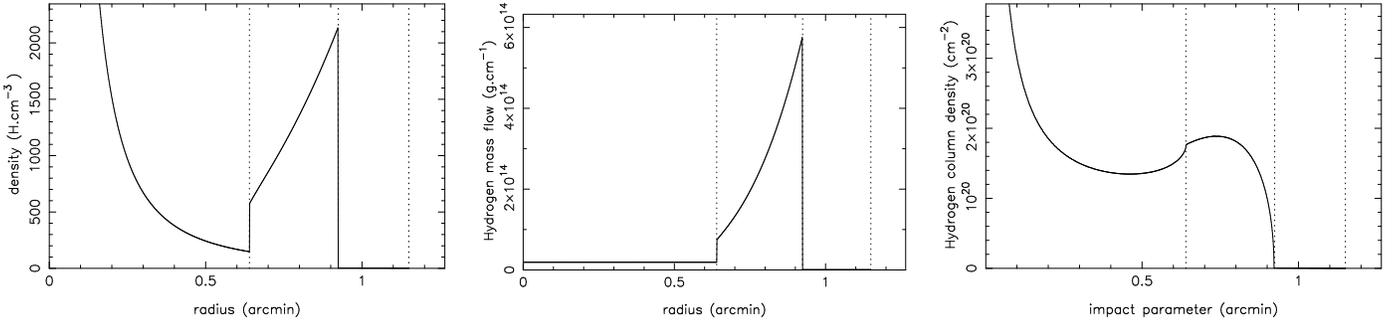

\centering
\begin{tabular}{c c c}
\includegraphics[width=4.2cm, angle=-90]{0089Fi8a.ps} &
\includegraphics[width=4.2cm, angle=-90]{0089Fi8b.ps} &
\includegraphics[width=4.2cm, angle=-90]{0089Fi8c.ps}
\end{tabular}
\caption{Left panel: atomic hydrogen density profile. Center panel: atomic 
	hydrogen mass-flow profile. Right panel: atomic hydrogen column density 
	calculated by the model. The vertical dotted lines show the radii, 
	$\rm r_1$, $\rm r_f$ and $\rm r_2$, used in the model.}
\label{profdMcol}
\end{figure*}

	As our model assumes spherical symmetry, we performed a fitting 
	on a symmetrized map, i.e. a map in which the offset positions 
	have been averaged (Fig.~\ref{model_HI}, upper panel). In the model, 
	the total flux is forced to be equal to that measured in the map 
	(i.e. 1.22 Jy $\times$ \kms). We set the central velocity at 28.8 \kms. 
	The results are summarized in Table~\ref{modelfit}.
	In this simulation, the temperature and the velocity are 
	constant inside r$_1$ (resp. 20 K and 4.2~\kms ). The shock at r$_1$ 
	decreases the velocity (increases the density) by a factor of 3.9 and 
	the temperature rises to 530\,K (Figs.~\ref{profvt} \& \ref{profdMcol}). 
	Then, inside the region of compressed matter (between r$_1$ and 
	r$\rm _f$), the temperature 
	decreases to $\sim$\,175 K. The physical conditions between r$\rm _f$ 
	and r$_2$ are in fact not constrained either by our model or by the 
	data at 21 cm, and in Table~\ref{modelfit} they are only extrapolated 
	(for more details, see Libert et al. 2007).

\begin{table}
\caption{Model parameters (d = 137 pc), with notations the same as 
	in Libert et al. (2007)}
\begin{tabular}{ll}
\hline
$\dot{M}$ (in hydrogen)           & 1.27 $\times$ 10$^{-7}$ 
					M$_{\odot}$ yr$^{-1}$ \\
$\mu$                             & 1.3\\
t$_1$                             & 5\,927 years\\
t$_{DS}$                          & 36\,800 years\\
r$_1$                             & 2.55 $\times$ 10$^{-2}$ pc (0.64 $'$)\\
r$_f$                             & 3.67 $\times$ 10$^{-2}$ pc (0.92 $'$)\\
r$_2$                             & 4.58 $\times$ 10$^{-2}$ pc (1.15 $'$)\\
T$_0$($\equiv$ T$_1^-$), T$_1^+$  & 20 K, 528 K\\
T$_f$ (= T$_2$)                   & 175 K\\
v$_0$($\equiv$ v$_1^-$), v$_1^+$  & 4.2 \kms, 1.07 \kms\\
v$_f$                             & 0.16 \kms\\
v$_2$                             & 1.2 \kms\\
n$_1^-$, n$_1^+$                  & 148 H\,cm$^{-3}$, 578 H\,cm$^{-3}$\\
n$_f^-$, n$_f^+$                  & 2.1 $\times$ 10$^{3}$ H\,cm$^{-3}$, 
					2.5 H\,cm$^{-3}$\\
n$_2$                             & 1.3 H\,cm$^{-3}$\\
M$_{r < r_1}$ (in hydrogen)       & 0.75 10$^{-3}$ M$_{\odot}$ \\
M$_{DT,CS}$   (in hydrogen)       & 4.67 10$^{-3}$ M$_{\odot}$ \\
M$_{DT,EX}$   (in hydrogen)       & 0.010 10$^{-3}$ M$_{\odot}$ \\
\hline
\end{tabular}
\label{modelfit}
\end{table}

	The model assumes that the \HI emission is optically thin 
	($\tau$\,$\ll$\,1). This can be verified using the output column 
	density profile (Fig.~\ref{profdMcol}, right panel) 
	and the expression $\tau$\,$=$\,5.50\,$\times$\,$10^{-19}\frac{N_H}{T\Delta V}$ 
	(Eq.\,12, Libert et al., 2007) where N$_{\rm H}$ is in cm$^{-2}$ and 
	$\Delta$V, the line width, in~\kms . With T $>$ 175 K and $\Delta$V 
	$\sim$ 3.8~\kms , the optical depth stays below 0.5 at all impact 
	parameters $>$ 0.1$'$ from the central star. 

	In general, the model provides a satisfactory fit to the symmetrized 
	\HI spectra that 
	we have obtained on RX Lep. However, it predicts a flux above 
	the observations on the central 
	position and below at 22$'$ in declination. This can be 
	understood as a consequence of the 4.4$'$ offset to the south 
	noted in Sect.~\ref{HI}. By moving the model 4.4$'$ 
	south and 0.4$'$ west with respect to the central star, we can improve 
	the fit to the observed data (Fig.~\ref{model_HI}, lower panel). 
	This gives support to the offset values that we have determined by 
	Gaussian-fitting in Sect.~\ref{HI}. Yet, the spectrum on the position 
	at 22$'$ south is not well reproduced, suggesting that the source is 
	more extended along the north-south direction than along the east-west one, 
	as suspected in Sect.~\ref{HI}.

	In the past (Y CVn, Libert et al. 2007), we already attempted to 
	better fit the data by a shift in 
	position to take into account the deformation of the envelope by the ISM. 
	However, this approach is artificial because we used a spherical model 
	that is not centered on the star position. It is only meant to illustrate 
	the need for an \HI mapping of this interesting source with a better 
	spatial resolution and the need to develop a true non-spherical modeling 
	of detached shells.

\section{Discussion}
\label{discussions}
	RX Lep shows evidence of a circumstellar envelope of $\sim$\,0.01\,M$_{\odot}$ 
	that may be the result of its stellar wind decelerated 
	by the external medium. This star is an oxygen-rich, 
	semi-regular variable on the E-AGB (no evidence of Tc, cf. 
	Sect.~\ref{source}). It is in the same evolutionary stage as EP Aqr 
	and X Her, which have also been detected in H\,{\sc {i}} and for which 
	the emission at 21 cm shows evidence of significant circumstellar 
	envelopes (Le\,Bertre \& G\'{e}rard 2004; Gardan et al. 2006). 
	We note that these 3 stars share the same variability properties 
	and have about the same luminosity ($\sim$ 4500\,L$_{\odot}$) 
	and the same effective temperature ($\sim$ 3200 K). It suggests 
	that mass loss can already occur efficiently for this type of star 
	on the E-AGB.

	Our model strongly relies on the mass-loss rate estimated from CO 
	observations. It is noteworthy that this estimate is consistent with 
	Reimers' relation (Reimers 1978). Indeed, by adopting M $\sim$\,3 
	M$_{\odot}$, L $\sim$ 4500 L$_{\odot}$ and T$\rm _{eff}$ $\sim$ 3300 
	K (Sect.~\ref{source}), this relation gives $\rm \dot{M}$ $\sim$ 2 
	$\times$ 10$^{-7}$ M$_{\odot}$ yr$^{-1}$. However, the luminosity 
	was probably lower in the past, as was the mass-loss rate. This suggests 
	that the age (42\,700 years) is underestimated.
	
	The model and our observations together put constraints on the physical 
	conditions within the CS between the termination shock (r$_1$) 
	and the interface (r$_{\rm f}$). Directly from the observations, the mean 
	temperature should be $\lesssim$ 300 K. 
	Based on the assumption of an adiabatic shock at r$_1$, it implies an 
	increase in temperature to $\sim$ 500 K. Thus, the gas must be cooled 
	down in the CS. Estimating the cooling rate is difficult at such low 
	temperatures. Nevertheless, the \HI line-profiles put constraints 
	on the behavior of the temperature because it is coupled to the 
	kinematics (Libert et al. 2007). The temperature profile shown in 
	Fig.~\ref{profvt} (lower panel) yields the best fit to the shape of 
	the \HI spectra. Between r$_{\rm f}$ and r$_2$, our model has only 
	been extrapolated. This region is 
	probably dominated by interstellar material flowing at $\sim$\,50\,\kms~through 
	the bow shock. The material should be denser than assumed  
	in our model; indeed, this region is fed by the interstellar medium 
	that has been swept up through the relative motion of RX Lep 
	circumstellar shell, at $\sim$ 50 \kms, rather 
	than by the expansion of the shell during the same period of 4\,10$^4$ years. 
	Also it is expected to be ionized, and therefore might not contribute
	significantly to the \HI emission that we detected.
	
	The \HI data indicate that RX Lep's CS is offset about 4$'$ to the 
	south and 0.5$'$ to the west. This agrees within 25$^{\circ}$ with 
	the direction of the proper motion 
	given by Hipparcos (PA $\sim$\,31$^{\circ}$). In addition, the model 
	hints that the shell is not completely spherical, and that RX Lep is 
	slightly elongated mostly in the N/S direction. This suggests 
	that the elongated shape observed in \HI is connected to the motion 
	of RX Lep through the local ISM. Villaver et al (2003) have made 
	numerical simulations of the evolution of a low-mass star moving 
	supersonically through the ISM and find that, due to the ram-pressure 
	stripping, most of the mass ejected during the AGB phase is left downstream. 
	The left panel in their Fig. 1 shows that CSs are progressively distorted 
	and become elongated in the direction of the motion with respect to their 
	surrounding ISM. The 25$^{\circ}$ difference between 
	the space motion of the star and the elongation of the shell could 
	then be due to the intrinsic velocity of the ISM local to RX Lep 
	relative to the LSR. Such intrinsic motions are currently found in the local 
	solar neighborhood (Redfield \& Linsky, 2008, Fig. 16 and references therein). 
	A significant fraction of the velocity of the local ISM is a reflection 
	of the solar motion; nevertheless, the velocity of the Sun with respect to 
	the LSR (13.4 \kms, according to Dehnen \& Binney, 1998) is 25$^{\circ}$ away 
	from the direction of the velocity of the average local ISM with respect 
	to the Sun (26.7\,\kms).

	It is also worth noting that for RX Lep 
	there is no significant difference between the \HI and CO central 
	velocities, as if the interaction only occurs in the plane of the sky. 

\begin{figure}
\centering
\includegraphics[width=4.4cm]{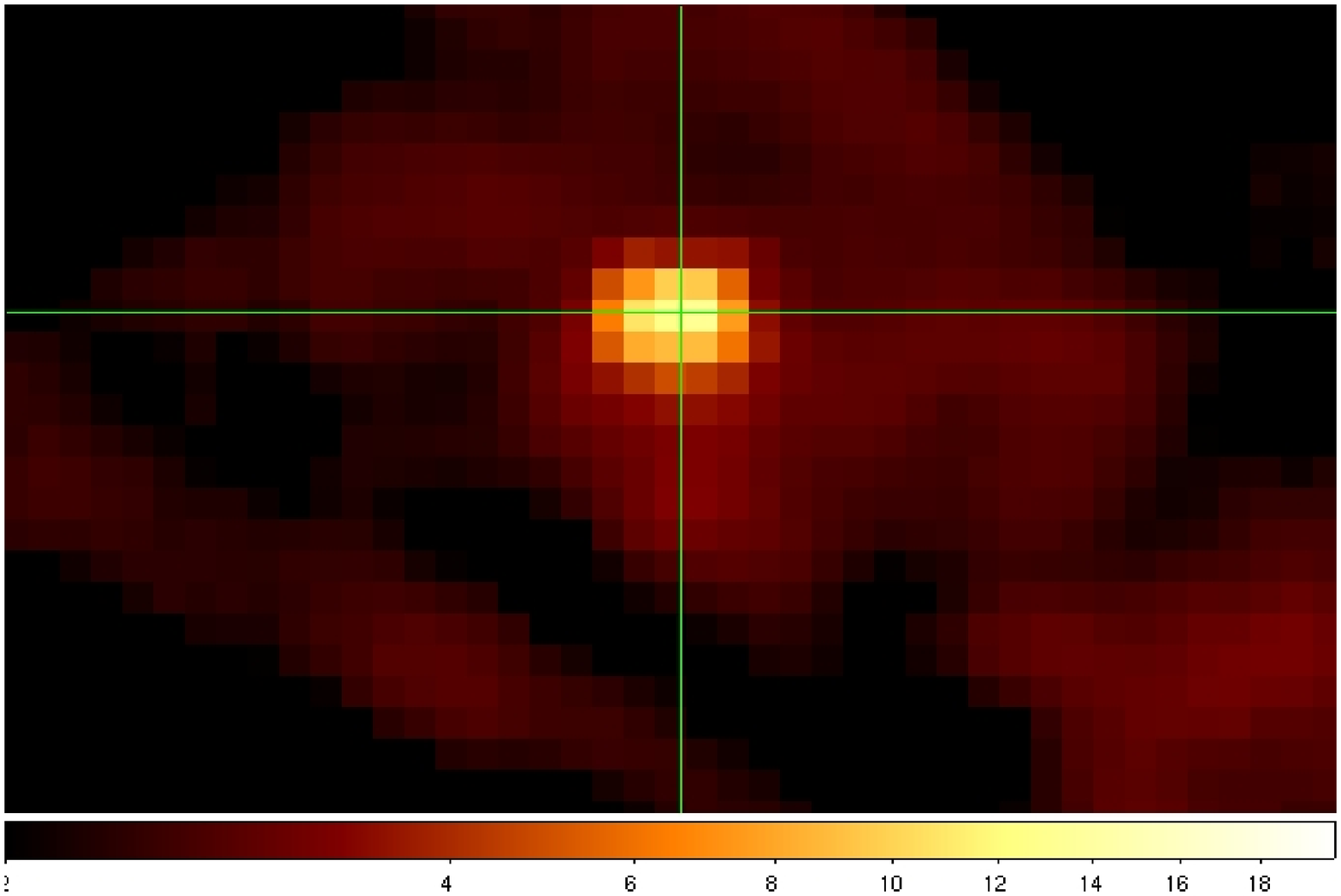} \includegraphics[width=4.4cm]{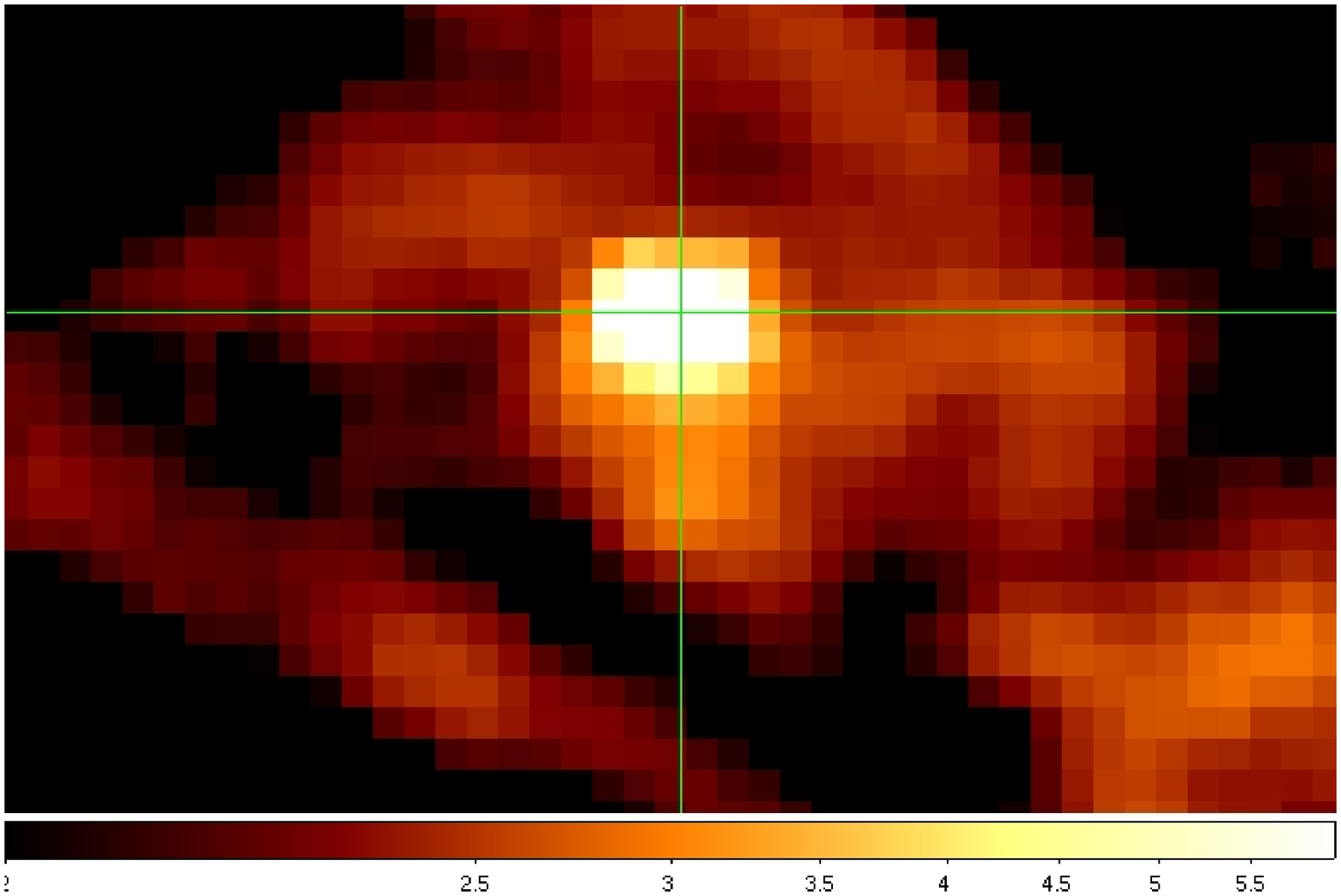}\\
\includegraphics[width=4.4cm]{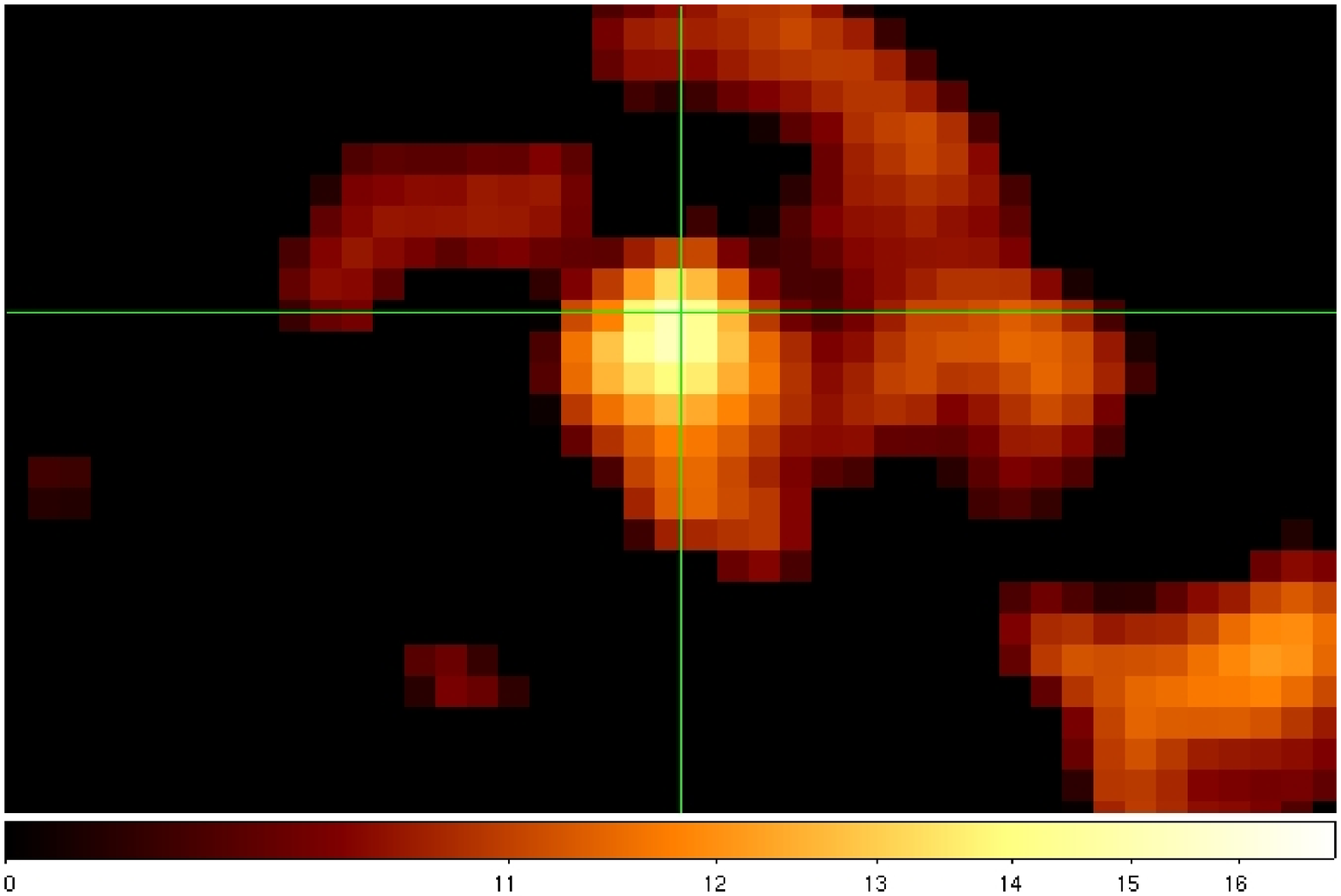} \includegraphics[width=4.4cm]{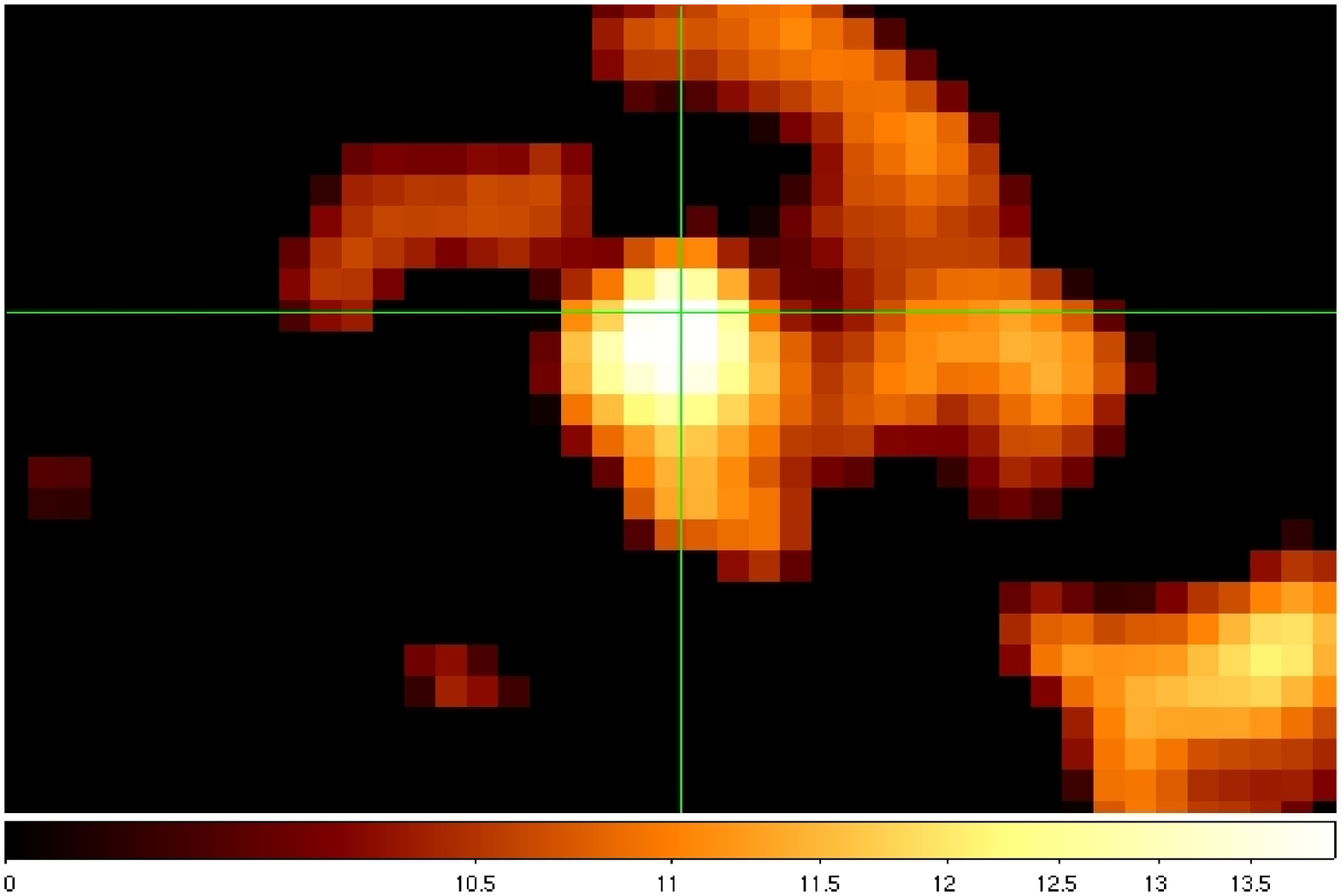}
\caption{Reprocessed IRAS images (IRIS) at 60 $\mu$m (upper panels) and 100 $\mu$m 
	(lower panels). To enhance the suspected extended emission to the south, 
	we present a non-saturated version (left) and a saturated one (right) for both 
	wavelengths. The field is $\sim$ 65$'$ $\times$ 39$'$ and the green 
	reticles mark the position of RX Lep 
	(north is to the top and east to the left).}
\label{IRIS}
\end{figure}

	We have examined the IRAS maps that have been reprocessed recently 
	by Miville-Desch\^{e}nes \& Lagache (2005; IRIS: Improved Reprocessing 
	of the IRAS Survey). The 60 $\mu$m and 100 $\mu$m images (Fig.~\ref{IRIS}) 
	suggest a small extended source ($\phi$ $\sim$ 6$'$ - 8$'$). The source 
	at 100 $\mu$m might be shifted by $\sim$ 2$'$ to the south. There 
	is also an extension ($\sim$ 12$'$) to the south; however, 
	it might be an artifact due to the satellite 
	scanning in the north-south direction. In these images, we cannot 
	confirm the X0509-119 offset with respect to RX Lep (cf 
	Sect.~\ref{source}). New data with a better spatial resolution, e.g. from 
	the Far Infrared All-Sky Survey of Akari, may help to clarify this 
	situation.

	In their \HI survey of evolved stars, G\'{e}rard \& Le Bertre (2006) 
	found that the line-profiles are Gaussian-shaped and often offset 
	with respect to the stellar velocity by $\sim$ 1-3 \kms~towards 0 \kms~
	LSR. Several \HI sources were also noted to be spatially offset from the 
	central star. They suggest that these effects could be related to 
	a non-isotropic interaction with the local ISM. Matthews \& Reid (2007) 
	have imaged the \HI emission around RS Cnc with the VLA. They find 
	that it is elongated with a peak on the stellar position and a filament 
	extending $\sim$\,6$'$ to the northwest, in a direction opposite 
	to that given by the proper motion. Recently, Matthews et al. (2008) have 
	imaged the \HI emission of Mira with the VLA. As for RS Cnc, they find 
	a ``head-tail'' morphology oriented along the star proper motion and 
	consistent, on large scales, with the far-ultraviolet emission discovered 
	by GALEX (Martin et al. 2007). Furthermore, the high spectral resolution 
	\HI data obtained with the NRT along the 2-degree GALEX trail reveal 
	a deceleration of the gas caused by interaction with the local ISM. 
	Finally, using Spitzer MIPS data obtained on R Hya at 70 $\mu$m, Ueta 
	et al. (2006) discovered a bow-shock structure \underline{ahead} of the 
	star in the direction of its motion. The excess emission that delineates 
	this bow shock is seen at 70 $\mu$m, but not at 160 $\mu$m; it may 
	partly come from the [O\,{\sc {i}}] cooling line at 63 $\mu$m. Although 
	we have presently no direct evidence in \HI of a bow-shock, both 
	structures, bow-shock and \HI trail, should be causally related (Wareing 
	et al. 2006). In fact, as the 
	velocity of these sources with respect to the ISM is 
	often high (see e.g. Nyman et al. 1992, or Mennessier et al. 2001), 
	the interstellar material is probably ionized through the bow shock, 
	so that we may never detect directly such a bow-shock structure in \HI 
	at 21 cm. Better tracers would likely be line emission in the 
	UV/optical/IR ranges (H$_{\alpha}$, [Fe\,{\sc ii}], [O\,{\sc i}], etc.).

	We therefore have a convergent set of results that shows that AGB stars 
	are associated with large-scale circumstellar shells distorted by the 
	motion of these evolved objects through the ISM (Villaver et al. 2003). 
	We suggest that RX Lep 
	is one more source in such a case. That the source is elongated 
	in the same direction as its offset and nearly opposite to the direction 
	of motion, argues in favor of a head-tail morphology. 

\section{Conclusions}

	We detected CO(2-1) and \HI line emissions from the semi-regular 
	oxygen-rich E-AGB star, RX Lep. These emissions indicate a stellar outflow 
	at a velocity $\sim$ 4.2 \kms~and a rate 
	$\sim$\,1.7\,$\times$\,10$^{-7}$\,M$_{\odot}$\,yr$^{-1}$, 
	with a duration of 4 $\times$ 10$^{4}$ years. The 
	\HI source has a size of $\sim$ 2$'$ ($\approx$ 0.08 pc) in the 
	east-west direction and possibly 15$'$ ($\approx$ 0.6 pc) in the 
	north-south direction.

	The modeling of the \HI line profiles obtained at different positions 
	suggests that the outflow is slowed down by the interaction with the 
	ambient ISM, and that 
	the external part of RX Lep circumstellar shell is made of compressed 
	material, at $\sim$ 200 K, as in the well-known detached shell around 
	Y CVn.

	The elongated shape of the RX Lep \HI source is compatible with the 
	direction of its proper motion, as in the cases of Mira and RS Cnc, 
	which have already been studied at high angular resolution with the VLA.

\begin{acknowledgements}
	The Nan\c{c}ay Radio Observatory is the Unit\'e scientifique de 
	Nan\c{c}ay of the Observatoire de Paris, associated as Unit\'e de 
	Service et de Recherche (USR) No. B704 to the French Centre National 
	de la Recherche Scientifique (CNRS). The Nan\c{c}ay Observatory also 
	gratefully acknowledges the financial support of the Conseil R\'egional 
	de la R\'egion Centre in France. 
	This research made use of the SIMBAD database, operated at the CDS, 
	Strasbourg, France, and of the NASA's Astrophysics Data System. We thank 
	the referee, Dr T. Ueta, and Dr L. Matthews for helpful suggestions.
\end{acknowledgements}

\end{document}